# A Novel Collaborative Cognitive Dynamic Network Architecture

Beatriz Lorenzo, F. Javier Gonzalez-Castano, Yuguang Fang

*Abstract*—Increasing mobile data demands in current cellular networks and proliferation of advanced handheld devices have given rise to a new generation of dynamic network architectures (DNAs). In a DNA, users share their connectivities and act as access points providing Internet connections for others without additional network infrastructure cost. A large number of users and their dynamic connections make DNA highly adaptive to variations in the network and suitable for low cost ubiquitous Internet connectivity. In this article, we propose a novel collaborative cognitive dynamic network architecture (CDNA) which incorporates cognitive capabilities to exploit underutilized spectrum in a more flexible and intelligent way. The design principles of CDNA are perfectly aligned to the functionality requirements of future 5G wireless networks such as energy and spectrum efficiency, scalability, dynamic reconfigurability, support for multi-hop communications, infrastructure sharing, and multi-operator cooperation. A case study with a new resource allocation problem enabled by CDNA is conducted using matching theory with pricing to illustrate the potential benefits of CDNA for users and operators, tackle user associations for data and spectrum trading with low complexity, and enable self-organizing capabilities. Finally, possible challenges and future research directions are given.

*Keywords*—Future network architecture, cognitive radio, network reconfigurability, matching theory, data and spectrum trading, QoS.

## I. INTRODUCTION

The rapid growth of wireless devices and services exacerbates the problem of spectrum scarcity and poses potential challenges for mobile operators, especially in terms of quality of service provisioning. According to Cisco [1], mobile traffic is expected to grow up to 1000 times by 2020, overwhelming the cellular infrastructure. The ever-increasing mobile applications such as mobile social networks, online gaming and high-definition video streaming may further accelerate this process.

To accommodate the explosive growth in mobile traffic and devices, a new paradigm shift is needed in the design of 5G network architectures, moving from infrastructure-centric networks with exclusive spectrum ownership towards dynamic user-centric approaches incorporating cognitive capabilities. In this context, a new generation of dynamic network architectures (DNAs) is emerging where users share their connections and act as access points for others without additional infrastructure cost. A framework for topology optimization in a DNA is developed in [2] by considering users' quality of service (QoS) requirements on access point selection. Flexible reconfigurability to adapt to traffic dynamics is achieved by using a genetic algorithm. Perez-Romero et al. [3] propose power-efficient resource allocation schemes for a DNA and show power reductions of approximately 40% with respect to the conventional cellular approach. Moreover, several business models and incentive mechanisms have been proposed and shown the potential of DNA to generate profits for users and operators [4], [5].

Considering that a large portion of spectrum is underutilized temporally and spatially, integrating cognitive radio capabilities into the DNA for spectrum harvesting [6] will facilitate access to additional unused spectrum to meet the growing spectrum demand and mitigate interference among adjacent DNAs. Furthermore, the harvested spectrum may have different propagation or penetration properties, which can be exploited to support diverse applications with various QoS requirements.

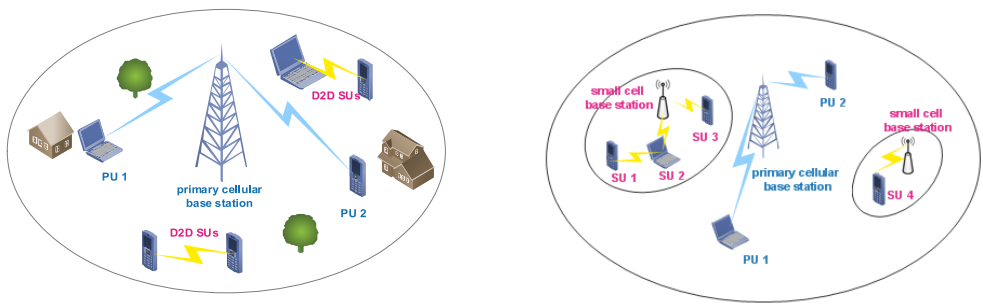

Fig. 1. Cognitive network architectures

Novel architectures for cognitive networks based on D2D [7], small cells [8] and multi-hop communications [6], [9] have been proposed to further increase spectrum efficiency. These architectures are illustrated in Fig. 1. D2D spectrum sharing intends to offload traffic from the cellular infrastructure when source and destination are close to each other. Interference minimization is a key issue when multiple D2D pairs share the same resources. In [7] the spectrum sharing problem between a set of D2D pairs and multiple co-located cellular networks is studied. Each D2D link can either access a sub-band occupied by a cellular subscriber or obtain an empty sub-band for its exclusive use. The problem is formulated as a Bayesian non-transferable utility overlapping coalition formation game.

Cognitive capabilities have been utilized to reduce intercell interference among the macrocell and small cells to deploy spectrum efficient cognitive heterogeneous networks [8]. Although the introduction of small access points can solve the capacity demand issue, the overall energy consumption and infrastructure cost are significantly increased. Several works propose to use sleep mode techniques [10] to overcome these problems. However, high traffic demand fluctuations over space, time, and frequency render these techniques infeasible. Enabling multi-hop communications in cognitive networks can further increase coverage and spectrum efficiency by exploiting locally available channels and support dynamic traffic distributions without additional infrastructure cost [6]. The importance of backup channels to increase link reliability and robustness in multi-hop cognitive networks is addressed in [11]. They show that high reliability requirements can be met when there is traffic misbalance between primary and secondary networks. However, the required backup channels and switching delay increase with the number of hops. M. Pan et al. [9] propose a spectrum trading system for multi-hop cognitive capacity harvesting networks (CCHNs) where partial infrastructure such as cognitive radio routers have to be deployed. They present a theoretical study on the optimal session based spectrum trading problem under uncertain spectrum supply and multiple cross-layer constraints. Their discussion on related works is a good survey on spectrum trading.

In this article, we leverage the potentials of DNAs and cognitive networks and present a novel collaborative cognitive dynamic network architecture (CDNA) which integrates cognitive radio capabilities into the DNA concept for efficient spectrum sharing across the network. The proposed CDNA encourages user level collaboration to relieve congestion on the secondary network by offloading traffic through available PUs. This enables efficient use of the available connectivities and spectrum by reusing local available channels at different locations without requiring additional infrastructure. In addition, the high density of users provide multiple options for connectivity and reconfigurability to traffic dynamics and have the potential to generate benefits for both primary and secondary networks. In the sequel, we describe the architecture and the parties involved in details. After that, we discuss its numerous possibilities in the context of 5G. Then, a case study to illustrate its performance is provided together with an analytical framework to efficiently tackle resource allocation problems. Some possible challenges and future research directions are also outlined. Finally, our conclusions are presented.

## II. A COLLABORATIVE COGNITIVE DYNAMIC NETWORK ARCHITECTURE

In this section, we present a new architecture which incorporates cognitive capabilities into the DNA concept, resulting into a Cognitive Dynamic Network Architecture (CDNA). In CDNA, primary users (PUs) share their connectivities and act as access points for secondary users (SUs) for some rewards. Here, data is forwarded to a BS by a PU on the primary network which reduces the reliability concern of multi-hop transmissions to the first hop (SU-PU link). As shown in Fig. 2, the proposed architecture involves the following parties: primary operator (PO), secondary operator (SO) and their end users. SO has its own spectrum bands, although potentially congested already, and cannot satisfy the quality of service (QoS) requirements for its SUs. Based on the SUs demand, the SO negotiates with the PO a service-level agreement (SLA) for the connection: resources (the number of needed PUs, channels, data) and the price. The PO will select a set of PUs and encourage them with a reward to share their connectivities with the SUs. Then, SUs will associate with the available PUs for data transmissions. In what follows, we elaborate the roles and interactions of the parties involved.

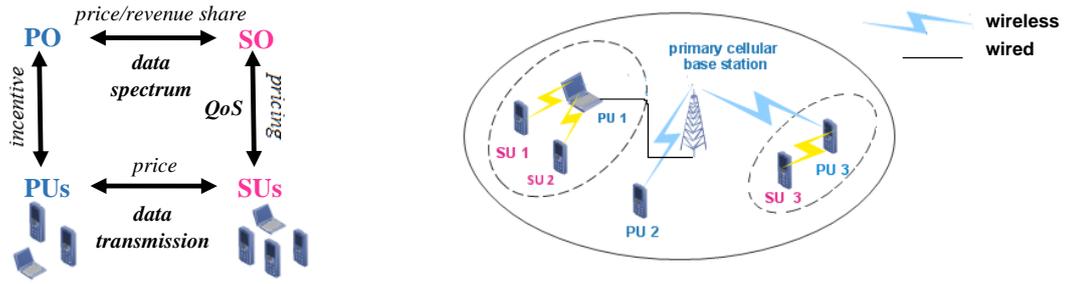

Fig. 2. Cognitive Dynamic Network Architecture
a) Interaction between the parties involved.  b) PU-SU association.

*A. PO and SO*

The PO leases its idle licensed spectrum bands to the SO to obtain a profit. There are two options for the contract: a) the PO shares with the SO the spectrum concurrently, which results in mutual interference between PUs and SUs. In this case, the PO charges for this interference and the SO pays the corresponding interference cost; b) the PO shares with the SO the spectrum in an orthogonal manner which eliminates the interference between SUs and PUs, and the SO pays the cost for the transmission opportunities. The PO and the SO also negotiate the price for the number of PUs and the amount of data to be forwarded by the PUs. Auctions are widely used to negotiate prices of resources between PO and SO. Other options include, revenue sharing techniques, price differentiation, which considers the budget of the SUs, and contract based approaches. The dynamics of CDNA as result of changes in PUs availability and SUs demand calls for flexible contracts between PO and SO. For instance, a two-level agreement with an initial fee for establishing cooperation and a termination fee for the data forwarded will provide flexible price adjustment to network dynamics. Advanced SU devices with sensing capabilities can opportunistically use the temporally available spectrum to transmit to PUs, but have to stop using it when a PU returns to the channel. Given the uncertainty on the channel availability, in this case the spectrum will be sold at a lower price.

*B. SO and SU*

The SO harvests licensed spectrum bands, purchases spectrum bands at different locations and performs channel allocation for SU links. The amount of SUs serviced by a SO depends on the remaining spectrum not being used by the PUs. Besides, the SO has knowledge of SUs connectivity requirements which may include connectivity duration, reliability, delay or throughput. The SO will allocate the channels and available PUs to the SUs to maximize the social welfare. Therefore, the resources will be assigned to satisfy the QoS requirements for the SUs that value them the highest. The SUs can have guaranteed access to the resources for the time duration of the contract between the SO and PO. If the contract stipulates that SUs and PUs can share the spectrum in a concurrent way, then the SO will keep track of the availability of PUs and channels to mitigate interference. Additionally, the SO can exploit its cognitive capabilities to facilitate seamless mobility to SUs by effectively using dynamic spectrum access and hence, reduce handoffs. The handoff here is a negotiation to get connectivity through an available PU and access to the local available channels. Indeed, the dynamic adaptation to the changes of PUs and channels availability significantly impacts the performance of this architecture.

*C. SU and PU*

Once the PO and the SO have reached a collaboration agreement, PUs and SUs will associate for data transmissions. The association can be performed centrally by the SO or distributely, letting SUs and PUs associate based on their connectivity requirements. We anticipate that centralized solutions will entail high complexity due to the high density of wireless networks and will incur high management latency. In contrast, distributed solutions favor scalable and flexible reconfiguration of the network. More details on the complexity of resource management schemes are provided in Section IV. PUs can benefit directly from data forwarding by setting a reasonable price considering the competition between primary users in the data market. Sophisticated SU terminals with sensing capabilities could also detect the available spectrum holes and choose the most suitable PU that provides the best utility and price ratio. The PO and SO could benefit as well from the distributed implementation and earn additional benefits directly from PUs and SUs.

*D. PO and PU*

The success of this architecture depends on PUs active participation to share their resources. Hence, the PO will encourage its PUs to share their resources (available data, power) in exchange for specific incentives such as extension of their data plan or economic incentives, letting them earn some percentage of the negotiated price for the data transmission. The PUs contribute to extend the coverage of the primary network and efficiently reuse the available data and spectrum. In return, the PO will share the benefits obtained from this architecture with the PUs.

Apart from stimulating PUs' participation, the PO is in charge of the following tasks:

- Supervise the transactions between SUs and PUs to guarantee the fair trade of the negotiated resources and avoid any misbehaviors that may harm the continuation of this service.
- Resource management to keep track of available resources at PUs, perform resource allocation to PUs and provide fair compensation for the shared resources.
- Mobility management and localization to track the positions of PUs and provide the most suitable PUs to reconfigure the network in a time efficient manner.
- Interoperability to ensure that devices can be reconfigured to connect to CDNA.

## III. ADVANTAGES OF CDNA

The design principles of CDNA are perfectly aligned to the functionality requirements of future 5G wireless networks (i.e., energy and spectrum efficiency, scalability, reconfigurability, support for heterogeneous QoS, infrastructure sharing, multi-hop transmission and multi-operator cooperation).

### A. Energy Efficiency

The network infrastructure of CDNA is built on demand, which is an energy efficient approach since there is no energy cost on maintaining additional infrastructure. The PUs will act as access points whenever it is agreed and there is a SU demanding their services. Providing incentives to PUs to share their connectivity brings a cost to the network. This cost depends on their remaining battery power and data. Hence, the PO may design incentive mechanisms to stimulate the least costly PUs to participate in order to satisfy the traffic demand. As more PUs agree to share their connectivities, the coverage of CDNA increases and thus, it becomes attractive to more SUs to adopt their service. However, as more SUs join CDNA, the competition for connectivity increases and they may encounter increasingly congested access to PUs. Therefore, the PO will increase or decrease the incentive offered to PUs to control the performance.

The cognitive capabilities of CDNA also contribute to its energy efficient operations by minimizing waste due to interference. Specifically, spectrum sensing enables interference avoidance between SUs and protection of PUs.

### B. Spectrum Efficiency

The cognitive capabilities of CDNA and the availability of PUs at different locations enable efficient reuse of the spectrum between SUs. The same spectrum band can be assigned to different SUs located far away from each other as the interference will be low. Besides, CDNA provides efficient spectrum sharing across the network between PUs and SUs by exploiting the available channels locally and temporally due to the traffic dynamics in both networks.

### C. Network Reconfigurability

The traffic fluctuations in both primary and secondary networks creates opportunities for cooperation. The congested network (secondary network) will negotiate agreements for resource leasing with the least congested primary network. As already mentioned, the PO is responsible for the interoperability of the devices to facilitate the cooperation. The changes in the availability of PU and SU demands require CDNA to handle dynamic reconfiguration to maintain the expected performance. Once the cooperation is established, the high density of wireless devices provides multiple options for connectivity which allows dynamic reconfigurability of CDNA over space, time, and frequency. Hence, CDNA exploits the density and dynamics of wireless networks to provide scalable and flexible reconfiguration of the network.

### D. Support for Heterogeneous QoS Requirements

CDNA provides support for heterogeneous QoS requirements given the heterogeneity of its components which include:

- wireless devices with different capabilities (laptops, smart phones, tablets),
- different mobility patterns (fixed, mobile),
- coexistence of wired and wireless Internet with heterogeneous access points (BS, WLAN AP, femtocell,…) and,
- heterogeneity of spectrum bands.

For instance, a PU can share its Internet connection on a laptop and provide high data rate connection to wired Internet or Wi-Fi access points. Fixed PU access points are also desirable to establish long range links for highly mobile SUs to avoid frequent disconnections. On the other hand, mobile PU access points such as smart phones or tablets can provide wireless access for delay tolerant applications as the mobility of both PUs and SUs may incur frequent disconnections. Short transmission distances and short rendezvous between PUs and SUs are favorable for high date rate transmissions. In addition, the cognitive capabilities of CDNA enable to switch to high spectrum bands with better penetration for indoor environments (e.g., TV bands) or low spectrum bands for long transmission distances.

Apart from the previously described two-hop transmissions, CDNA may support longer multi-hop transmissions to extend coverage of wireless networks. Here, the multi-hop route would be built on demand without additional infrastructure cost [16]. A route discovery protocol is needed to find the most suitable routes according to the SUs demands.

### E. Multi-operator Cooperation

The variations in spectrum usage, channel quality and coverage in different operators' networks generate plenty of cooperation opportunities between operators, which CDNA can exploit to improve network performance. As result of this cooperation, CDNA increases the coverage of the network and the likelihood of overlaps with other operator networks with different hops, allowing users to choose among multiple access opportunities.

Most existing works on multi-operator cooperation focus on spectrum sharing at the system level [6]. Thus, the traffic variations at individual cells are not considered, which limits the gains achieved with such schemes. Just recently, multi-operator cooperation has been addressed in the context of cellular networks and the observed benefits have attracted great interest [12]. The idea is to let each participating operator to contribute some spectrum in the spectrum pool to better handle traffic variations. Then, the gains achieved by traffic offloading from a cellular network to a WLAN are quantified by an equivalent increase in opportunistic capacity proportional to the ratio of aggregate coverage of cellular networks and WLANs. Similar analysis can be conducted for multi-operator cooperation in CDNA focusing on infrastructure sharing and data trading among others. This analysis will open new research lines on the macro-economic aspects of multi-operator cooperation in CDNA.

## IV. CASE STUDY: DATA AND SPECTRUM TRADING IN CDNA

### A. Matching Theory Based Framework

In this section, we conduct a case study to show the performance of CDNA. We focus on data and spectrum trading in CDNA which involves SU-PU association, channel allocation and price determination. To fully exploit the potentials of this architecture, we are interested in developing distributed mechanisms to solve the previous problem which can track network dynamics with low communications overhead. The dynamics here result from varying resource demand from SUs and resource supply due to PU activity and channel availability. The efficiency of matching theory and its suitability for cognitive networks has been corroborated by recent works such as [13], [14]. The resource allocation in cognitive networks involves two set of agents (PUs and SUs) which are matched based on their preferences. Through sequences of acceptances and rejections, a stable matching is reached, and coordination can be achieved in a distributed way with low communication overhead.

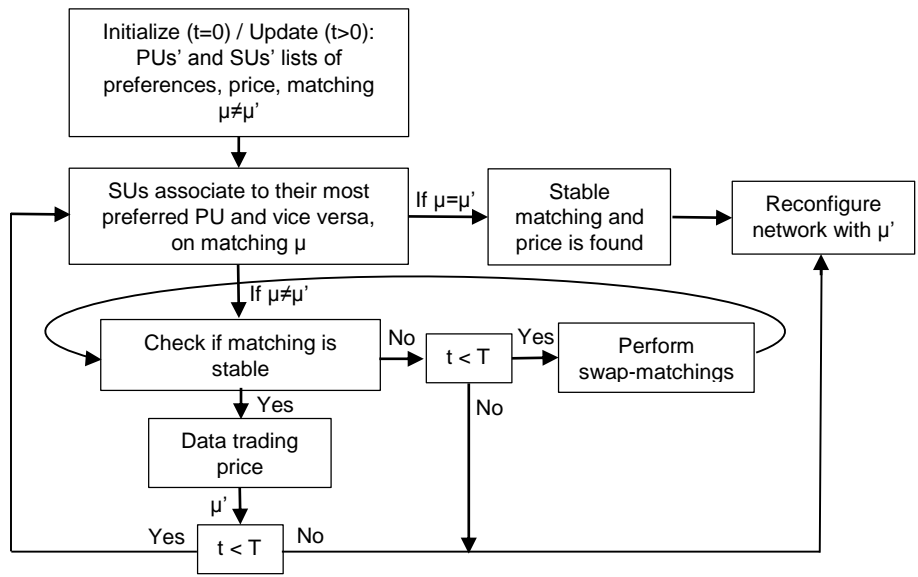

Fig. 3. Distributed matching with pricing algorithm.

In this article, we develop an algorithm based on many-to-one matching with pricing as shown in Fig. 3 and we solve the matching for a snapshot of the network traffic (duration *T*). For any other traffic situation in the primary or secondary network, the same procedure applies. Here, multiple SUs are assigned to a PU that satisfies their QoS requirements as long as the data quota of that PU is not exceeded. SUs' preferences over PUs capture the throughput, connectivity duration and price for the data transmitted, while PUs' preferences capture the data volume sold and price trade-off that they can achieve. The SO provides the information regarding channel availability to SUs. The data transmitted varies with channel availability, so each set of users builds a list with the preferences over the other set per channel. To efficiently use the spectrum, we assume that channels are reused by SUs at different locations. Therefore, the preferences of SUs and PUs are influenced by the existing matching and the proposed game can be classified as a matching game with externalities. In fact, for a SU-PU link, the available data volume depends on the data volume requested by other buyers. Due to these externalities, the traditional solutions based on the deferred acceptance algorithm in [13] are unsuitable. Instead, our algorithm considers a new stability concept, based on the idea of swap-matching [15] and extends it to define *SU-PU-swap* and *channel-swap* in our data and spectrum trading market. As users measure externalities, they modify their preferences and change their choices by performing swap-matchings to other SU-PU associations or channel allocations as shown in Fig. 3.

Once a stable SU-PU association and channel allocation is found, the SU and PU negotiate the price for the resources. The current demand of the SU and supply of the PU are dynamically adjusted in the association process until the price for the resources converges to the stable price −following the principle of market equilibrium. Hence, if the data demand for a particular PU is high, the price will increase, which favors load balancing between PUs. Otherwise, it will decrease.

### B. Performance Evaluation

We consider a snapshot of the network traffic in a CDNA with 10 PUs and 20 SUs, each with a monthly contract for a data volume of 10 GB. A SU seeks connectivity through a PU in the CDNA when the SO cannot satisfy its QoS requirements. There are two reasons for this situation: the SU has run out of its data plan or the SO is congested. In the former case, the SU either pays an overage fee for extra data to transmit in the secondary network or transmit in the CDNA (using the data plan of the PU) paying the data cost imposed by the PU. Whereas in the latter case, the SO offers a reward to the SU to transmit in the CDNA and contribute to the resolution of the network congestion. We set all users' transmit powers to 20 mW, the noise level to $10^{-7}$ W/Hz and the path loss exponent to $\alpha = 3$. The minimum SINR requirement varies in [1, 20] dB and the duration of connectivity varies in [0, 15] minutes. The energy consumption is set to 0.257 J/MB, the price of a 10GB data plan to 10€, and the price per GB traded in CDNA between 0.1 and 1€ [5]. Figure 4 shows the average utility per SU, where the probability of exceeding the data plan is $e = 0.8$. The results are presented with respect to the transmission range, which varies from 20 to 200 m, which corresponds to the minimum SINR from 20 dB to 1dB in the same range. The performance is compared with random matching and worst-case utility. The latter refers to the matching that provides the lowest SU utility. We can see that, as the transmission range increases and with it the connection alternatives, our proposed scheme yields a performance advantage of 25% in terms of utility improvement relative to random matching and 50% to the worst-case utility. Similar gains were observed in the PU utility.

Figure 5 shows the average number of SUs transmitting to their secondary macro-cell BS (SBS), as the number of SUs increases for a constant number of PUs. For high *e* values, the number of users transmitting to SBS decreases and SUs prefer to join the CDNA to purchase additional data. Furthermore, if the overage price $p$ in the secondary network increases with respect to the price $\pi$ in CDNA, an additional decrease in number of users is observed in the secondary macro-cell. In particular, when price $p$ doubles, the number of SUs transmitting to SBS decreases by about 25%.

Figure 6 shows the revenue of the operator obtained through CDNA and SBS operation. We can see that CDNA yields higher revenue to the operator than SBS when the SU has exceeded its data plan as a higher volume of data is transmitted through CDNA. The revenue reaches 200% when users have high QoS requirements (low transmission range).

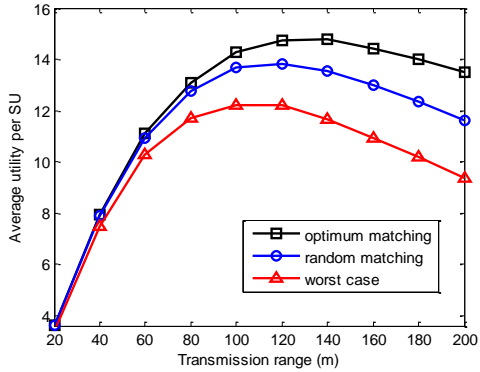

Fig. 4. Comparison of the average utility per SU.

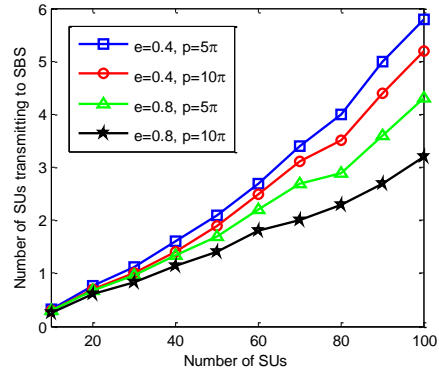

Fig. 5. Number of SUs tx. to SBS.

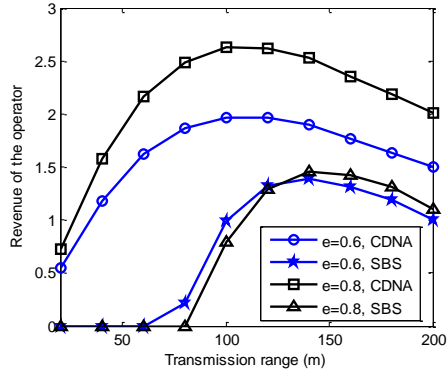

Fig. 6. Operator revenue in CDNA and SBS.

It was observed that the number of iterations needed for convergence of the algorithm increases up to 30 for medium-sized networks (less than 50 SUs), and remains constant for high-density networks (over 50 SUs). This is because, as the density increases, adjacent PUs offers similar performance levels and, thus, the number of swaps decreases. Hence, the algorithm presents a reasonable convergence time even in high dense networks which makes it suitable to track network dynamics [2].

Many extensions to the previous framework can be envisioned. For example, the reputation of PUs and SUs, and mobility patterns can be incorporated in the preference functions. More sophisticated pricing mechanisms can be designed for the negotiation between the SO and the PO to control the traffic in the secondary network and CDNA.

In cognitive radio networks, centralized optimization solutions are undesirable since PUs and SUs belong to different operators and cannot be centrally controlled. Besides, the proposed resource allocation problem formulated as a centralized optimization yields a combinatorial problem which is NP-hard with complexity of the order $\mathcal{O}(N^{MB})$ limiting its applicability in a dynamic network. A non-cooperative game is also applicable here, although it will have more limitations that include the need to observe all players' preferences and the fact that the solution would not account for two-sided stability as previously explained.

The matching theory algorithm provides self-organizing solutions which are highly desirable in CDNA due to network density and scale. In the worst case, when all SUs have the same preference ordering per PU and channel, the overall number of SU proposals (messages exchanged between SUs and PUs) is of the order $\mathcal{O}(NMB)$. In each iteration of the matching algorithm, the PU broadcasts the channels used in the current matching so that the remaining SUs can estimate the quality of service before making a new proposal to the PU. Moreover, the algorithm can be terminated at any time associated with a desirable complexity level. Such properties of the algorithm make it scalable and flexible to changes in the network and also to specific complexity or implementation requirements.

## V. Challenges and Future Research

The CDNA opens multiple future research directions related to the following challenges.

### A. Reconfigurable User Equipment

In the proposed network architecture, the PO and the SO provide information regarding the available channels. Nevertheless, embedded sensing and reconfiguration abilities in mobile devices will be important in the future for better spectral efficiency. In IEEE 802.22, a variety of radio technologies are being studied. It intends to enable spectrum sensing for wireless regional area networks (WRANs). In addition, various cognitive radio platforms support reconfigurability, such as Universal Software Radio Peripheral (USRP), USRP2, Wireless Open Access Research Platform (WARP), and Wireless Network Centric Cognitive Radio (WiNC2R).

Terminal reconfigurability facilitates spectrum handoff and enables dynamic adaptation between channels coexisting locally with primary users, and improves interference avoidance by sensing locally available channels.

### B. Economic Models

As CDNA has the potential to generate benefits for both users and operators, we divide the economic models into micro- and macro-economic models:

- *Micro-economics* studies the benefits that CDNA brings to users such as incentives received from the operator to stimulate user participation. These incentives are related to the amount of data shared and the remaining battery power at the terminal. Monetary incentives and extra data quota are the most common incentives. Nevertheless, operators may encourage the autonomy of users by letting them trade with their data plans and further profit from the transaction. The pricing mechanism should provide fair trade and resource allocation. Thus, pricing should capture the heterogeneous connectivity requirements from SUs such as throughput, connectivity duration and budget, as well as the available resources and channel availability at the PUs. The operator will set up the conditions for the trade and may benefit as well from the transaction. As connectivity in CDNA is enabled by users, incentive mechanisms that take into account human psychology deserve further study.

- *Macro-economics* studies the benefits achieved from multi-operator cooperation. As already mentioned, these benefits result from spectrum sharing, data trading, and energy and infrastructure savings. The PO and SO can share the benefits obtained in CDNA by using a revenue sharing scheme or agree in a long term cooperation. As multiple parties are involved in the trade, the negotiation requires different levels which can be modeled as Stackelberg or iterative games.

C. *Security Design*

Security measures at user and operator level are needed to guarantee the well-functioning of this architecture.

- *User level*: in CDNA users rely on others to transmit their data but not all users are reliable. For instance, users may lie about their availability and receive a reward but, later on, disrupt the transmission. Both PUs and SUs can take security measures such as installing trusted software on their devices. This security investment reduces the probability that users would be vulnerable to an attack and also the probability that other users are under attack when acting as access points. The security level should be available in the association process between SUs and PUs. If the security level provided by the PU is higher than the security level of the SU requesting the connection, the PU will charge an extra cost to compensate for scanning the system or take extra security measures after serving the SU.

- *Operator level:* the PO should supervise the transactions and guarantee that the amount of data traded and the price are as agreed in the negotiation process. Besides, the SO can build up a trust management scheme and store the reputation of the PUs based on the previous experience of other SUs.

## CONCLUSIONS

In this article, we have proposed a new cognitive dynamic network architecture (CDNA) for future 5G wireless networks which leverages the density and heterogeneity of wireless devices to provide ubiquitous Internet connectivity. The advantages of this architecture include energy and spectrum efficiency, dynamic reconfigurability, support for heterogeneous connectivity requirements and a new business opportunities. To fully exploit its potential, a distributed matching algorithm is developed for resource allocation, which allows the users to self-organize into a stable matching for data and spectrum trading. The algorithm can track network dynamics with low communication overhead which facilitates scalable and flexible reconfiguration of CDNA due to changes in spectrum and user activity.

This article aims to articulate the potential of users' active role in future wireless networks in improving network performance and inspire research on such a paradigm shift towards collaborative architectures. A number of interesting directions for future research are also highlighted.